\documentclass[draft]{agujournal}

\draftfalse

\journalname{JGR-Space Physics}

\usepackage{array}
\usepackage{makecell}
\usepackage{datetime}
\usepackage{ragged2e}

\begin{document}
\justify
%
%



\title{On the response of Martian ionosphere to the passage of a corotating 
interaction region: MAVEN observations}

%
%




\authors{C. Krishnaprasad\affil{1}, Smitha V. Thampi\affil{1}, and Anil Bhardwaj\affil{2}}


\affiliation{1}{Space Physics Laboratory, Vikram Sarabhai Space Centre, Thiruvananthapuram 695022, 
India}
\affiliation{2}{Physical Research Laboratory, Ahmedabad 380009, India}





\correspondingauthor{C. Krishnaprasad}{kpchirakkil@gmail.com}




\begin{keypoints}
\item First direct comparison of the response of dayside and nightside ionosphere of Mars to the 
impact of a corotating interaction region (CIR)

\item Observations of enhanced flux of precipitating suprathermal ($>$25 eV) heavy ions in the 
Martian exosphere during CIR period

\item Indication of enhanced pickup ion induced depletion of nightside ionosphere during CIR event
\end{keypoints}


%
%


\begin{abstract}
The response of Martian ionosphere to the passage of Corotating Interaction Region (CIR) of June 
2015 is studied using observations from several instruments aboard the Mars Atmosphere and Volatile 
EvolutioN (MAVEN) mission. An intense CIR arrived at Mars on 22 June 2015, during which the upstream 
solar wind and interplanetary conditions were monitored by the Solar Wind Ion Analyzer, Solar Wind 
Electron Analyzer, Magnetometer, and Solar Energetic Particle instruments aboard MAVEN. The CIR 
event was characterized by enhancements in solar wind density, velocity, and dynamic pressure, and 
increased \& fluctuating interplanetary magnetic field, and was accompanied by enhanced fluxes of 
solar energetic particles. The Langmuir Probe and Waves (LPW) instrument onboard MAVEN provided the 
ionospheric observations such as electron density and electron temperature during this period. The 
dayside ionosphere is significantly compressed only near the peak of solar wind dynamic pressure 
enhancement ($\sim$14 nPa). In contrast, on the nightside, the electron density remains depleted for 
a longer period of time. The electron temperatures are also enhanced during the period of electron 
depletion on the nightside. The STATIC (Suprathermal and Thermal Ion Composition) measurements show 
enhanced fluxes of suprathermal heavy ions in the Martian exosphere during CIR period, and evidences 
for enhanced tailward flow of these pickup ions. The analysis suggests that the nightside ionosphere 
is primarily controlled by the precipitating solar energetic particles and pickup ions transported 
across the Martian terminator, and depletes significantly when the heavy ion flux in the exosphere 
enhances. 
\end{abstract}

%
%

%


%
%
%
%

\section{Introduction}
The interaction of solar wind with Mars is different from that of Earth. This is primarily because 
of the absence of an intrinsic global dipolar magnetic field on Mars \citep{Russell1979, 
Acuna1998, Acuna1999}. A shock boundary defines the bow shock, 
where the solar wind changes from supersonic to subsonic speeds. The magnetosheath 
is a region of shocked, turbulent solar wind plasma behind the bow shock \citep{Lundin1991, 
Dubinin1997}. The induced magnetospheric boundary or magnetic pileup boundary formed from 
the interplanetary magnetic field (through induced currents) is another plasma boundary 
within the magnetosheath, where the induced magnetosphere divert a significant fraction of solar wind 
plasma from further penetration into the Martian upper atmosphere. Hence, magnetic pileup region 
is a transition region from solar wind plasma to planetary plasma \citep{Lillis2015}. 
Another plasma boundary called ionopause defines the altitude where the solar wind dynamic 
pressure balances the ionospheric pressure \citep{Duru2009, Vogt2015}. The topside ionosphere of 
Mars is mainly controlled from top by solar wind interaction \citep{Vogt2015}. The presence of remnant crustal magnetic fields (strongest in 
the southern hemisphere) makes the interaction picture more complex \citep{ArkaniHamed2004, Acuna1998, Connerney2005, 
Connerney2015, Lillis2015}.\\ 

The spatial variability in the coronal expansion and solar rotation cause solar wind flows 
of different speeds to become radially aligned. Compressive interaction regions are formed when a 
high\textendash speed solar wind runs into slower plasma ahead. These compression regions 
(bounded by the forward and reverse shocks) form spirals in the 
solar equatorial plane that corotate with the Sun. 
Because the pattern of compression rotates with the Sun when the outflow pattern from the Sun is 
time\textendash 
stationary, these high pressure regions are known as corotating interaction regions, or 
CIRs \citep{Parker1958, Sarabhai1963, Carovillano1969, 
Smith1976, Gosling1999}. 
The high\textendash speed streams (HSS) in CIRs are associated with 
coronal holes, which recur for more than one solar rotation. The track of interplanetary 
traveling CIRs from Sun through the interplanetary space to large 
heliospheric distances were previously 
studied by \citet{Williams2011} and \citet{Prise2015}, with observations from spacecrafts 
en\textendash route and in orbit around different solar system objects. The effect of CIRs on 
terrestrial ionosphere have been extensively studied 
(e.g. \citet{Borovsky2006}). But, a similar level of understanding is not achieved  on 
our understanding of the impacts of CIRs on Mars and Venus  because of the dearth of observations especially due to the lack of 
simultaneous solar wind and magnetic field monitors near the space environment of these planets.

Interplanetary coronal mass ejections (CMEs) and CIRs are the major solar wind dynamic pressure 
pulse events. These solar phenomena play a major role in space weather and ion escape at Mars 
\citep{Ma2014, Curry2015, Jakosky2015b, Sanchezcano2017}. Previously \citet{Ma2017} used a 
time-dependent global magnetohydrodynamic (MHD) model to investigate the response of the Martian 
ionosphere and induced 
magnetosphere to a large solar wind disturbance associated with the interplanetary CME on 8 March 
2015. The MAVEN observations as well as model results showed that ion escape rates could be an order 
of magnitude enhanced in response to the high solar wind dynamic pressure during the CME event. 
\citet{Luhmann2017} used a data-validated MHD model to study the same event and an extreme CME event 
of July 2012. Their results also suggest enhanced solar wind pressure, magnetic field, and 
convection electric field combine to produce strong magnetospheric coupling with important 
consequences in ionosphere energization and escape.\\

\citet{Hara2011} (\citet{Dieval2013}) first reported enhancement (reduction) in the precipitation of 
heavy ions (solar wind protons and alpha particles) due to finite gyroradius effects. The upstream 
solar wind dynamic pressure and interplanetary 
magnetic field (IMF) are important factors in controlling
the global spatial pattern and flux of ions precipitating into the Martian upper atmosphere, with 
intense ion precipitations when gyroradii of pickup ions (planetary 
ions accelerated by the motional solar wind convective electric field) are relatively small 
\citep{Hara2017}. The 
study by \citet{Martinez2019} showed increase in precipitation ion flux by more than one order of 
magnitude during the arrival of September 2017 CME event compared to average flux during quiet solar 
wind conditions. Heavy pickup ion precipitation is the primary cause of atmospheric sputtering 
\citep{Leblanc2015, Leblanc2018, Wang2015, Chaufray2007}.\\     

There are a few studies on the influence of CIRs and HSS on the Martian magnetosheath and 
ionosphere, primarily using data from plasma analyzers. For instance, \citet{Dubinin2009} studied 
the impact of February 2008 ionospheric storm, induced by a corotating interaction region, on the 
Martian topside ionosphere, using observations from Analyzer of Space Plasma and Energetic Atoms 
(ASPERA-3) and Mars Advanced Radar for Subsurface and Ionospheric Sounding (MARSIS) onboard Mars 
Express (MEX) spacecraft. They have found that the clouds of dense and high pressure solar wind 
plasma penetrate the induced magnetospheric boundary to lower altitudes. These solar wind plasma 
clouds scavenge the topside ionospheric plasma, making the latter become rarefied and fragmentary. 
This causes energization of ions and an enhanced loss of volatiles from Mars (a factor of $\geq$10 
enhancement in topside ionospheric erosion) \citep{Dubinin2009}. \citet{Morgan2010} have studied the 
CIR encounter with Mars on December 2015 using MARSIS/MEX radar sounding observations. They observed 
two radar absorption events separated by 26 days; and concluded that these surface reflection 
absorption events are caused by enhanced ionospheric ionization from high fluxes of energetic 
particles accelerated by the shocks associated 
with CIR. The Martian ionospheric variability during variable solar wind conditions, such as 
CIRs and CMEs were investigated by \citet{Opgenoorth2013}. The solar wind conditions were obtained 
from proxy measurements at 1 AU. The study reports magnetosphere and ionosphere compression during 
solar wind dynamic pressure variations, and  signatures of increased plasma transport over
the terminator and enhanced ion outflow from the upper atmosphere \citep{Opgenoorth2013, 
Sanchezcano2017}.\\

A characteristic heavy-ion signature was observed with ASPERA-3/MEX in the vicinity of 
Martian ionosphere during the passage of a CIR in September to October 
2007 \citep{Hara2011}. \citet{Wei2012} have observed enhanced escape flux of oxygen 
ions in the Martian magnetosphere using observations from ASPERA-3/MEX. They have also compared the 
O$^+$ escape from polar region of 
Earth with that of MEX observations for the same CIR event, and concluded that same level of 
increase in upwelling oxygen ions was observed on Earth's poles in comparison with Mars. The solar 
wind dynamic pressure enhancements was more affecting the ion escape on Mars, while dipole 
field effectively prevents such coupling of solar wind kinetic energy to planetary ions in the case 
of Earth \citep{Wei2012}. \citet{Elliott2013} have found that Martian magnetosheath electron fluxes 
are enhanced during the CIRs and HSS. They also concluded 
that the electron flux in 
the ionosphere of Mars does not respond to the CIRs and HSS, although the average 
electron energy is enhanced, as observed in the electron spectrometer in ASPERA-3/MEX. 
\citet{Harada2017} reported irregular structures in the topside Martian 
ionosphere following a CIR-related interplanetary shock using MARSIS/MEX and upstream MAVEN 
observations.\\

From the above discussion, it is clear that we need to improve further our understanding on the 
impact of CIRs on the Martian plasma environment, especially the ionosphere and the exosphere. 
Corotating solar wind streams of June 2015 provide a unique opportunity to study the response of 
Martian ionosphere to the CIR streams, using the suite of in\textendash situ particles and fields 
observations from Mars Atmosphere and Volatile EvolutioN (MAVEN \citep{Jakosky2015a}) spacecraft. 
This event is also unique as this is one of the largest CIR event, in terms of the solar wind 
dynamic pressure and density, ever observed at Mars and one of the highest recorded by MAVEN 
\citep{Lee2017}. We use observations from Langmuir Probe and Waves (LPW \citep{Andersson2015}) 
instrument to show the electron density as well as electron temperature measurements made by MAVEN 
during the period of study. We also use observations from Solar Wind Ion Analyzer (SWIA 
\citep{Halekas2015}), Magnetometer (MAG \citep{Connerney2015}), and Solar Energetic Particle (SEP 
\citep{Larson2015}) instruments to show the upstream solar wind, energetic particle, and magnetic 
field conditions, and Suprathermal and Thermal Ion Composition (STATIC \citep{McFadden2015}) 
instrument observations for the mass resolved  ion energy flux information in the lower energy range 
compared to SEP. The solar wind electron and ionospheric photoelectron energy flux observations from 
Solar Wind Electron Analyzer (SWEA \citep{Mitchell2016}) are also obtained during this period. 
Together, this is a unique and comprehensive dataset, covering a wide energy range, which can help 
to gain new insights into the impact of CIRs on Martian space weather. 

\section{Data and Method of Analysis}

The upstream solar wind parameters (density, dynamic pressure, and velocity) and IMF are computed 
from the SWIA onboard moments and MAG measurements (intensity and direction of the magnetic field) 
following the method by \citet{Halekas2016}. The SEP fluxes are obtained from the SEP instrument. 
The Level 2, Version 01, Revision 01 (V01\_R01) data of SWIA, Level 2, Version 01, Revision 02 
(V01\_R02) data of MAG, and Level 2, Version 04, Revision 02 (V04\_R02) data of SEP are used for 
upstream solar wind observations. The Level 2, Version 01, Revision 01 (V01\_R01) onboard survey 
spectral data from SWIA and Level 2, Version 04, Revision 01 (V04\_R01) survey spectral data from 
SWEA are used.\\

The LPW instrument provides observations of electron density (n$_e$), electron temperature (T$_e$), 
and electric field waves in the ionosphere of Mars \citep{Ergun2015, Fowler2015, Andrews2015}. 
There are two Langmuir probes, mounted on 7.1 m length booms, onboard MAVEN. It works in both LP 
mode and waves mode. The electron density and temperature in the LP mode are derived from the 
current\textendash voltage (I\textendash V) characteristics \citep{Andersson2015}. To understand the 
average quiet time picture of the ionospheric electron density profiles, the average of 6 orbits in 
the similar solar zenith angle (SZA) regime, prior to the CIR arrival are used.  
The LP mode electron temperature Level 2, Version 03, Revision 02 (V03\_R02) data and waves mode 
electron density Level 2, Version 02, Revision 02 (V02\_R02) data are used for analysis. The Level 
2, Version 02, Revision 00 (V02\_R00) ion energy flux (data product C0), and mass resolved ion 
energy flux (data product C6) from STATIC instrument are used for observations of energy spectrum of 
protons and heavy ions in the Martian ionosphere and exosphere \citep{Steckiewicz2015}. STATIC is an 
energy, mass, and angular ion spectrometer (consisting of a toroidal ``top hat'' electrostatic 
analyzer with a 360$^o\times$90$^o$ field-of-view, 
combined with a time-of-flight (TOF) velocity analyzer with 22.5$^o$ resolution in the detection 
plane) that can record ion fluxes, with a base time resolution of 4 
seconds, as a function of energy (0.1 eV - 30 keV), mass (1 - 70 amu), azimuth direction (0 - 
360$^o$), and deflection angle ($\sim\pm$45$^o$) \citep{McFadden2015}. The MAVEN datasets shown in 
this paper are latest available and are downloaded from the Planetary Data System 
(https://pds.nasa.gov/).\\

SWIA measures the energy and angular distributions of solar wind and magnetosheath ions of energy 
between 25 eV to 25 keV in 48 energy steps. SWEA measures the energy and angular distributions of 
solar wind and magnetosheath electrons and ionospheric photoelectrons of energy between 3 eV and 
4600 eV in 64 energy steps. SEP measures the energy spectrum and angular distribution of solar 
energetic ions of energy between 20 keV to 6 MeV in 28 energy steps and solar energetic electrons of 
energy between 20 keV to 1 MeV in 15 energy steps. STATIC measures the velocity distributions and 
mass composition of suprathermal and thermal ions in the energy range 0.1 eV to 30 keV and in the 
mass range 1 amu to 70 amu (64 energy bins and 2 mass bins in the data product C0, 32 energy bins 
and 64 mass bins in the data product C6) \citep{Jakosky2015a}. The Wang-Sheeley-Arge 
(WSA)\textendash ENLIL+Cone model simulations during June 2015 are taken from 
{http://helioweather.net/}, {https://iswa.ccmc.gsfc.nasa.gov/}. In this global heliospheric model, 
the solar coronal model WSA is coupled with the three-dimensional MHD numerical 
model ENLIL \citep{Odstrcil2003}, which is combined with the Cone model, so as to numerically 
simulate the interplanetary solar wind plasma and magnetic field conditions and to provide a global 
heliospheric context. The National Solar Observatory Global Oscillation Network Group (GONG) 
synoptic magnetograms are used as input for model runs (see \citet{Mays2015} and references therein 
for a detailed description of WSA\textendash ENLIL+Cone model).

\section{Observations}

The Solar Dynamic Observatory Atmospheric Imaging Assembly (SDO/AIA) composite 
images of the Earth-facing disk of the solar corona during May through July 2015 indicate the 
presence of persistent mid- to low-latitude coronal holes for several solar rotations 
\citep{Lee2017}. The images of coronal hole in May, June, and July are shown in Figure 7 of 
\citet{Lee2017}. The CIR associated with the coronal hole triggered a moderate geomagnetic storm on 
7 June 2015, and a minor geomagnetic storm on 4 July 2015 at Earth \citep{Lee2017}. The coronal hole 
rotated towards Mars, and the SWIA solar wind observations indicate the arrival of a CIR at Mars on 
22 June 2015. The WSA\textendash ENLIL+Cone simulations confirms that the same coronal hole with 
negative IMF polarity is rotated towards Mars and the stream arrive at Mars on 22 June 2015 (not 
illustrated, see \citet{Lee2017}, Figure 4d). Figure 1a\textendash1d shows the solar wind 
density, dynamic pressure, velocity, and 
IMF ($|$B$|$, B$_x$, B$_y$, and B$_z$) conditions during 16 to 24 June 2015. After a period of 
$\sim$26 days (that is, one solar rotation period), on 17 July 2015, the CIR recurred at Mars, but 
the event was weaker ($\sim$1 nPa enhancement in solar wind dynamic pressure) in comparison with the 
first one \citep{Lee2017} and therefore no significant effects were observed on the ionosphere. In 
this study, we present the observations only for the June 2015 event.

\begin{figure}
\begin{center}
\includegraphics[scale=0.65]{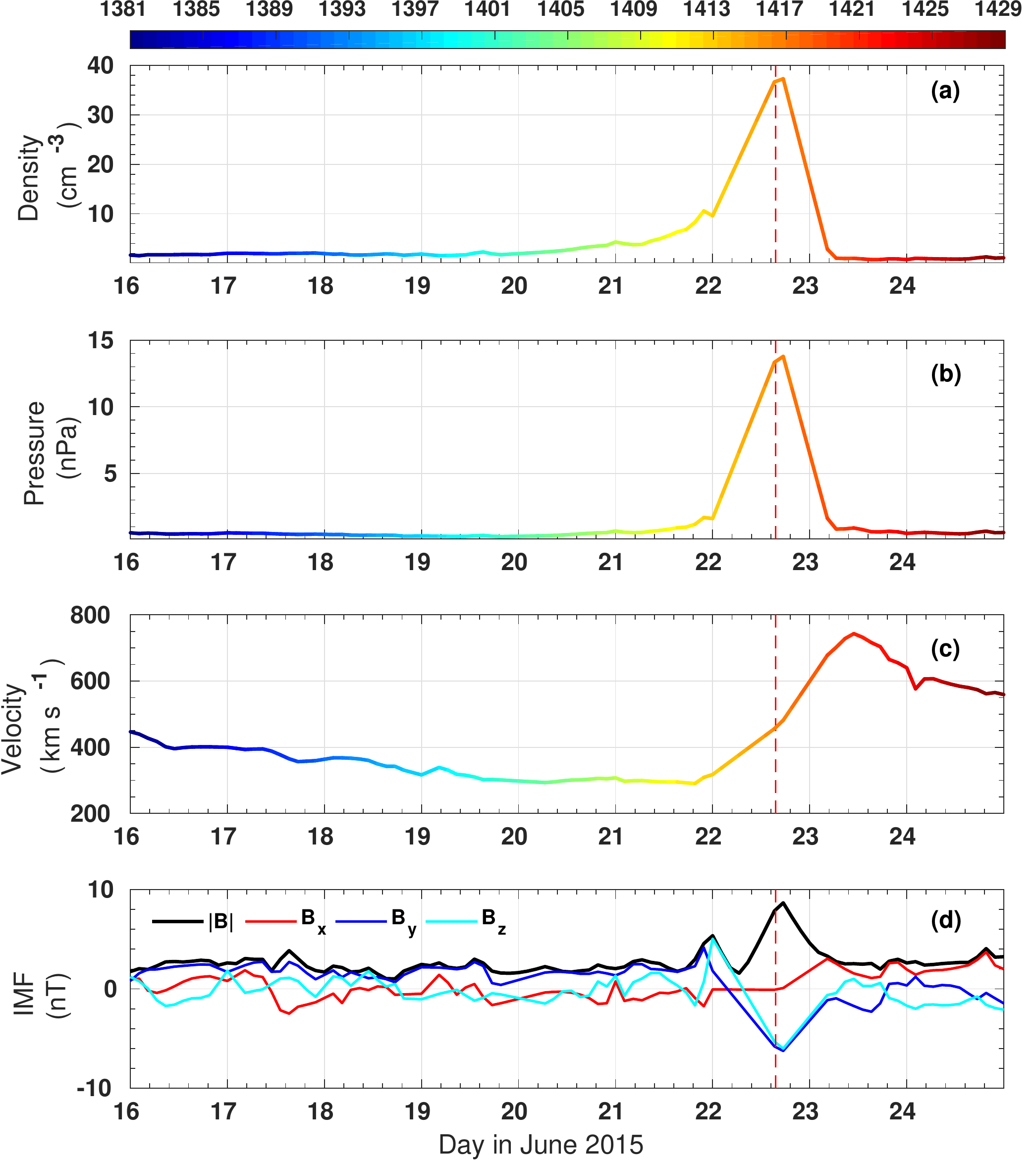}
\caption{The upstream solar wind and interplanetary magnetic field conditions during June 2015 
observed by SWIA: (a) solar wind density, (b) solar wind dynamic pressure, (c) solar wind 
velocity, and MAG: (d) IMF ($|$B$|$, B$_x$, B$_y$, and B$_z$). The 
color bar shows the orbits during the period. The periapsis of orbit 1416 inbound/1417 outbound is 
marked with 
red dashed line on the graph.}
\end{center}
\end{figure}

\begin{center}
\begin{figure}
\hspace{-4.5cm}
\includegraphics[width=1.6\linewidth]{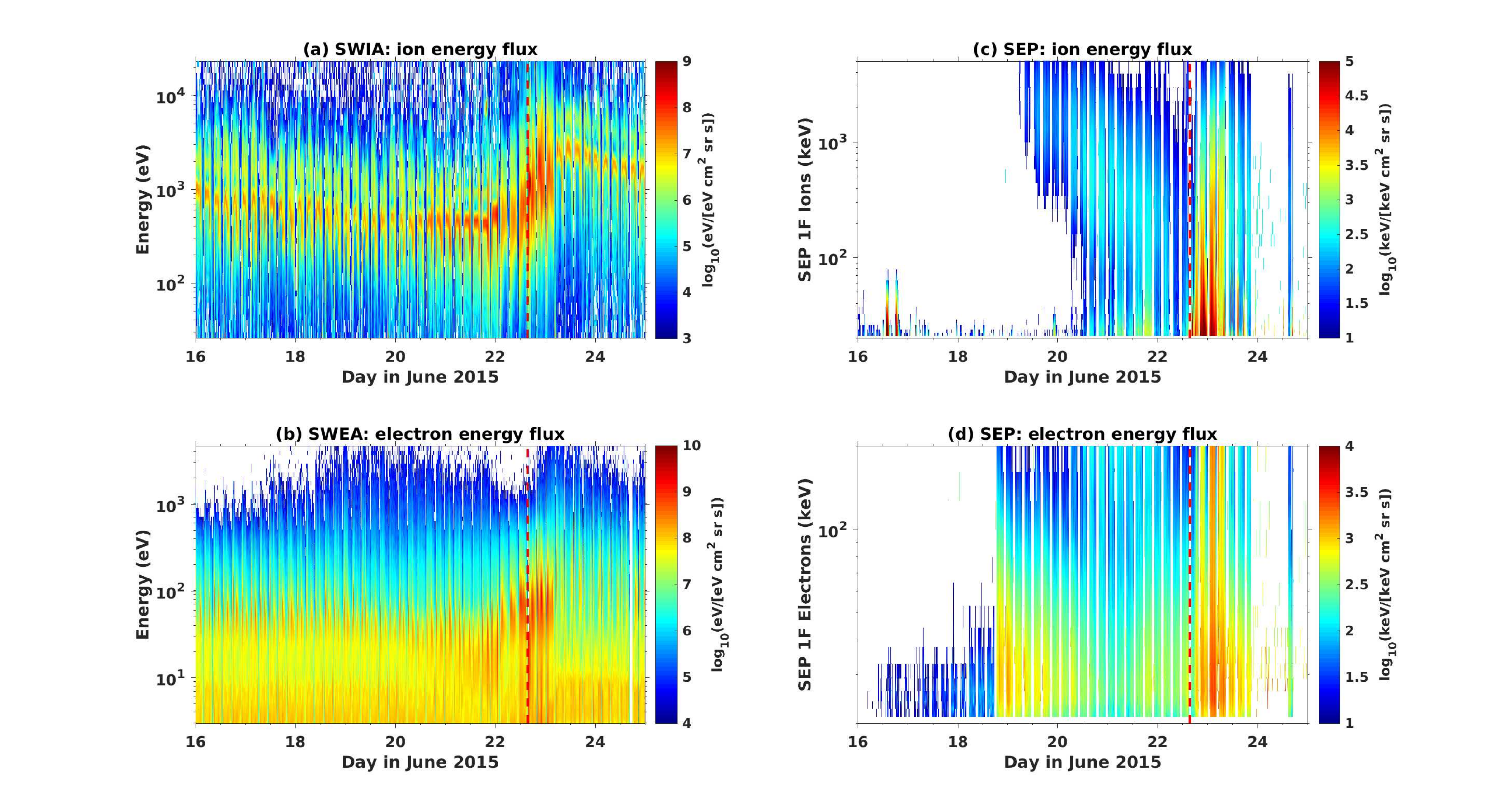}
\caption{(a) SWIA energy-time spectrogram of ion energy flux during 16 to 24 June 
2015. (b) SWEA energy-time spectrogram of electron energy flux during 16 to 24 
June 2015. (c) Differential energy flux of SEP 1F ions during 16 to 24 June 2015. (d) 
Differential energy flux of SEP 1F electrons during 16 to 24 June 2015. The periapsis of orbit 
1416 inbound/1417 outbound is marked with red dashed line on each spectrum. The initial white gaps 
in (c, d) are when the energy flux is below 10 keV/[keV cm$^2$ sr s], while white 
gaps on and after 24 June is due to absence of good quality data.}
\end{figure}
\end{center}

Figure 1 shows the  upstream solar wind and IMF conditions (in the right-handed Mars-Sun-Orbit [MSO] 
frame, with x-axis pointing toward the Sun and z-axis parallel to the normal to Mars' orbital plane) 
during 16-24 June 2015. The peak enhancement in dynamic pressure was $\sim$14 nPa, on 22 June 2015 
with peak density of $\sim$40 cm$^{-3}$, velocity of $\sim$800 km sec$^{-1}$ (which remains high, 
that is about $>$400 km sec$^{-1}$ until the end of the month), and total IMF peaked at $\sim$9 nT. 
Also the magnetic sector switches from ($-B_x$,$+B_y$) to ($+B_x$,$-B_y$), and $B_z$ oscillates from 
positive to negative to positive (Figure 1d). \citet{Lee2017} observed that the IMF configuration 
fluctuated between Parker spiral configuration and radial configuration, upon crossing the 
heliospheric current sheet associated with the HSS. The peak enhancements in upstream solar wind 
parameters was observed on 22 June 2015, 16:45 UTC when the compression region (peak density, 
dynamic pressure, and magnetic field) arrived at $\sim$1.5 
AU. These observations indicate that the CIR event was one of the  strongest observed stream 
interaction region event at Mars, since in most of the previously reported events the maximum solar 
wind dynamic pressure encountered was $\sim$10 nPa \citep{Dubinin2009}. This is more than the 
peak dynamic pressure observed by MAVEN during the 8 March 2015 interplanetary CME event 
\citep{Thampi2018, Lee2017}. The MAVEN orbit during the peak of CIR dynamic pressure enhancement was 
orbit 1416 (inbound)/1417 (outbound). (For the MAVEN data presented here, a new orbit 
number starts when the instrument is at the geometric periapsis, and is the same for the outbound 
and the next inbound sectors, and incremented at the next periapsis.)

Figure 2a shows the SWIA energy\textendash time spectrogram of omnidirectional ion 
energy flux during 16 to 24 June 2015. The solar wind ion energy corresponding to the peak flux 
increased from $\sim$500 eV 
(on 22 June 2015) to $\sim$5000 eV (on 23 June 2015). The solar wind ion energy 
spectrum follows the solar wind velocity pattern (Figure 1c). The solar 
wind ion flux enhanced from $\sim10^7$ to $>10^8$ eV/[eV cm$^2$ sr s] (with ion energies $>$ 3000 
eV) during the arrival of compression region. Figure 2b shows the SWEA 
energy\textendash time spectrogram 
of omnidirectional electron energy flux during 16 to 24 June 2015. An 
enhancement in energy of the peak flux (300 eV) and flux of electrons (from $\sim10^8$ to $>10^9$ 
eV/[eV cm$^2$ sr s]) during 22-23 June 2015 is observed.\\

The CIR event was associated with enhanced solar energetic particle (SEP) fluxes, with the arrival 
of SEP ions from 22 June 2015, 18:00 UTC (Figure 2c) and SEP electrons from 21 June 2015, 14:00 UTC 
(Figure 2d). The SEP ion enhancement is studied in detail in comparison with the observations at 1 
AU to understand the 
acceleration of particles associated with CIR events (\citet{Thampi2019}, manuscript submitted to 
ApJL). The  
SEP electron flux increased from $\sim10^2$ to $>10^3$ keV/[keV cm$^2$ sr s] with a corresponding 
order of magnitude increase in energy. Similarly, SEP ion flux increased to more than 
$10^3$ keV/[keV cm$^2$ sr s] in the higher energy channels. There is a 
significant enhancement in SEP electrons from 18 June, 19:00 UTC and SEP ions from 19 June, 15:00 
UTC prior to the arrival of the shock associated with CIR. These may be the SEPs associated with a 
CME that erupted on 18 June at $\sim$17:10 UTC, and are streamed along magnetic field 
lines towards Mars, even-though the CME shock did not hit Mars \citep{Lee2017}.\\

Table 1 (Table 2) shows the LPW periapsis measurement altitudes, longitudes, latitudes, solar zenith 
angles (SZAs), and time in UTC, with day of June 2015 and orbit number, for representative quiet 
orbits, 
viz, 1381, 1382, 1384, 1385, and 1386 and disturbed orbits, viz, 1415 (1416), 1416 (1417), 
1417 (1418), 1418 (1419), 1419 (1420), and 1420 (1421) during inbound (outbound) leg of MAVEN. (As 
mentioned earlier, the convention followed here is that a new orbit number starts when 
MAVEN is at the periapsis, and is the same for the outbound and the next inbound sectors, and 
incremented at the next periapsis.) The SZA on the topside night time ionosphere are between 
108$^\circ$ to 98$^\circ$ (inbound leg, altitude from 500 km to 200 km), while on the topside day 
time ionosphere are between 85$^\circ$ to 75$^\circ$ (outbound leg, altitude from 200 km to 500 
km), for the disturbed period orbits.

\begin{table}[ht]
\raggedright
\begin{tabular}{|c |c |c |c |c | c| c| c| c| c| c|}
\hline 
\thead{Day/} & \thead{UTC} & \thead{UTC} & \thead{Alt} & \thead{Alt} & \thead{Lon} & \thead{Lon} & 
\thead{Lat} & \thead{Lat} & \thead{SZA} & \thead{SZA} \\ 
\thead{Orbit \#} & \thead{(hr)} & \thead{(hr)} & \thead{(km)} & \thead{(km)} & \thead{(deg)} & 
\thead{(deg)} & 
\thead{(deg)} & \thead{(deg)} & \thead{(deg)} & \thead{(deg)} \\
\hline 
\thead{INBOUND} & From & To & From & To & From & To & From & To & From & To\\
\hline
16/1381&3.7&3.9&498.4&164.4&58.8&344.2&-74.3&-45.2&104.3&84.1\\
16/1382&8.2&8.4&498.9&164.9&124.5&49.5&-74.3&-45.3&104.5&84.4\\
16/1383&12.6&12.8&499.8&164&190.7&114.7&-74.3&-45.5&104.6&84.6\\
16/1384&17.1&17.3&499.6&164.8&256&180&-74.3&-45.7&104.7&84.8\\
16/1385&21.5&21.7&498.5&164.5&321.6&245.2&-74.2&-45.8&104.8&85.1\\
17/1386&2&2.2&498.4&163.7&27.4&310.4&-74.2&-45.9&104.9&85.3\\
\hline
22/1415&11&11.2&498.5&164.8&131.2&42.4&-73.3&-49.7&107.7&92\\
22/1416&15.4&15.6&498.9&163.9&197.2&107.7&-73.2&-49.8&107.8&92.2\\
22/1417&19.9&20.1&499.8&164.3&262.8&172.9&-73.1&-50&107.9&92.4\\
23/1418&0.3&0.5&498.9&164.3&328.3&238.2&-73.1&-50&107.9&92.6\\
23/1419&4.8&5&499.7&163.4&34.2&303.4&-73&-50.2&108&92.8\\
23/1420&9.2&9.4&499.6&164.5&99.6&8.7&-73&-50.3&108.1&93.1\\
\hline
\end{tabular}
\label{table:nonlin}
\bigskip
\caption{LPW measurement altitudes, longitudes, latitudes, SZAs, and time in UTC (with day of June 
2015 and orbit number) for representative quiet orbits 
and disturbed orbits during inbound leg of MAVEN (measurement below 500 km altitude).}
\end{table}


\begin{table}[h]
\raggedright
\begin{tabular}{|c |c |c |c |c | c| c| c| c| c| c|}
\hline 
\thead{Day/} & \thead{UTC} & \thead{UTC} & \thead{Alt} & \thead{Alt} & \thead{Lon} & 
\thead{Lon} & 
\thead{Lat} & \thead{Lat} & \thead{SZA} & \thead{SZA} \\ 
\thead{Orbit \#} & \thead{(hr)} & \thead{(hr)} & \thead{(km)} & \thead{(km)} & \thead{(deg)} & 
\thead{(deg)} & 
\thead{(deg)} & \thead{(deg)} & \thead{(deg)} & \thead{(deg)} \\
\hline
\thead{OUTBOUND} & From & To & From & To & From & To & From & To & From & To\\
\hline
15/1381&23.5&23.7&164&499.8&278.9&266.5&-44.9&-3.7&83.8&66.6\\
16/1382&3.9&4.1&164.4&499.6&344.2&331.7&-45.1&-3.8&84&66.9\\
16/1383&8.4&8.6&165&498.2&49.4&36.9&-45.2&-4.1&84.3&67.2\\
16/1384&12.8&13&164&499.2&114.7&102.1&-45.4&-4.1&84.5&67.4\\
16/1385&17.3&17.5&164.8&498.9&179.9&167.3&-45.5&-4.3&84.8&67.6\\
16/1386&21.7&21.9&164.5&499.2&245.2&232.5&-45.6&-4.4&85&67.9\\
\hline
22/1416&11.2&11.4&164.8&498.3&42.4&28.3&-49.6&-8.6&91.9&75.5\\
22/1417&15.6&15.8&163.9&499.9&107.6&93.5&-49.7&-8.6&92.1&75.7\\
22/1418&20.1&20.3&164.3&499.6&172.8&158.7&-49.8&-8.7&92.4&76\\
23/1419&0.5&0.7&164.3&498.6&238.1&223.9&-49.9&-8.9&92.6&76.3\\
23/1420&5&5.2&163.4&498.3&303.3&289.1&-50&-9&92.8&76.5\\
23/1421&9.4&9.6&164.5&499.9&8.6&354.3&-50.2&-9.2&93&76.7\\
\hline
\end{tabular}
\label{table:nonlin}
\bigskip
\caption{LPW measurement altitudes, longitudes, latitudes, SZAs, and time in UTC (with day of June 
2015 and orbit number) for representative quiet orbits and 
disturbed orbits during outbound leg of 
MAVEN (measurement below 500 km altitude).}
\end{table}

Figure 3a shows the nightside (near terminator) electron density profiles during 16 to 24 June 
2015. Orbits 1415 to 
1419 shows depletion in electron density compared to the mean quiet time profile. The mean quiet 
time profile with standard deviation is shown for comparison. The ionopause altitude observed 
for these two orbits are below 380 km. Also shown are two `normal' 
profiles. Here orbit 1384 is a pre-CIR nominal profile, while orbit 1420 is a post-CIR profile which has returned to 
the normal quiet time behavior, and both are within the standard deviation of the mean quiet time 
profile, confirming 
that they show typical quiet time state.\\

Figure 3b shows the dayside (near terminator) electron density profiles during 16 to 24 June 2015. 
Orbits 1417 and 1418 shows deviation in electron density profile compared to the mean quiet time 
profile. Here 
orbit 1385 is a pre-CIR nominal  profile, while orbit 1421 is a post-CIR profile, which has returned 
to the normal behavior. Here during orbit 1417, that is, the MAVEN orbit right during the peak of 
the solar wind dynamic pressure maximum, the electron density profile shows  a lower ionopause 
($\sim$400 km). But 
in contrast to the nightside, on the dayside the effects of CIR are observed in only one orbit, 
while on 
the nightside we observe the effect of CIR impact in five consecutive orbits. The orbit 
1418 passes through the crustal field region and is seen to be structured compared to the previous 
orbit profile \citep{Dong2015b}.\\

\begin{figure}
\begin{center}
\includegraphics[scale=0.9]{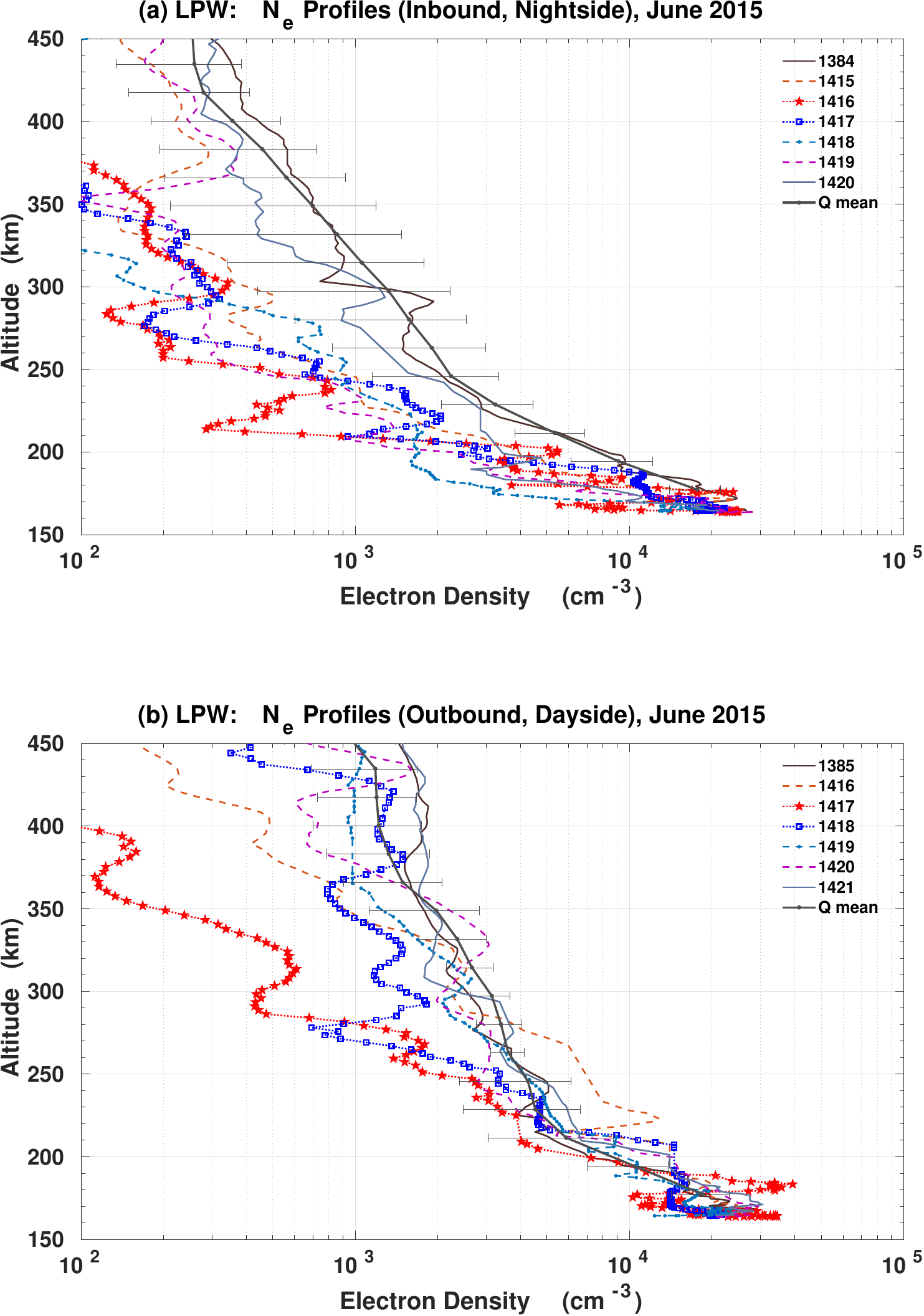}
\caption{(a) The nightside electron density profiles during June 2015 observed by LPW. The orbits 
1415, 1416, 1417, 
1418, \& 1419 are disturbed period orbits. Orbit 1384 is one of the pre-CIR quiet orbit and orbit 
1420 is post-CIR quiet orbit. The black thick solid line with errorbar represents the mean quiet 
time profile and standard deviation of quiet orbits 1381, 1382, 1383, 1384, 1385, \& 1386. (b) The 
dayside electron density profiles during June 2015 observed by LPW. The orbits 1416, 1417, 
1418, 1419, \& 1420 are disturbed period orbits. Orbit 1385 is one of the pre-CIR quiet orbit and 
orbit 1421 is post-CIR quiet orbit. The black thick solid line with errorbar represents the mean 
quiet time profile and standard deviation of quiet orbits 1381, 1382, 1383, 1384, 1385, \& 1386.}
\end{center}
\end{figure}

If we compare the quiet time behavior during day and night, it is evident that the variability 
is less in the dayside. The higher standard deviations for the nightside profile indicate that the 
nightside ionosphere of Mars is more variable compared to the dayside which corroborates with the 
previous observations \citep{Girazian2017}. The  profiles during the CIR event are significantly 
different from their 
representative quiet time mean profile and also well beyond the standard deviation, confirming 
that the deviations we discuss are significant and are beyond the expected quiet time orbit-to-orbit 
variations, both for the dayside and for the nightside.\\

\begin{figure}
\begin{center}
\includegraphics[scale=0.99]{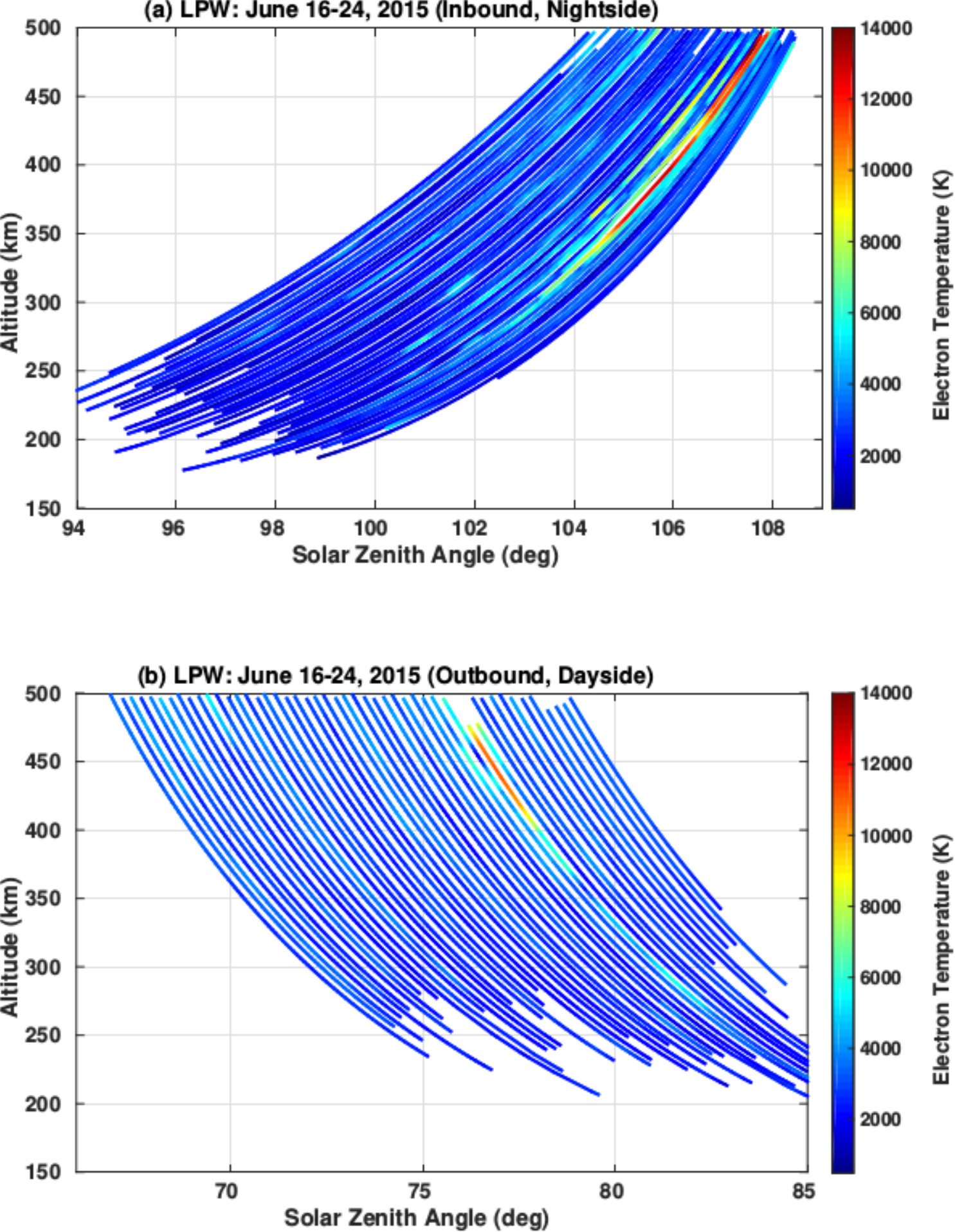} 
\caption{(a) The nightside electron temperature profiles with SZA during 16 to 24 June 2015 
observed by LPW (Orbits 1381 to 1428 -- from left to right). The nightside electron temperature is 
enhanced to more than 12000 K during inbound orbits 1416 and 1417. (b) The dayside electron 
temperature profiles with SZA during 16 to 24 June observed by LPW 2015 (Orbits 1382 to 1429 -- 
from left to right). The dayside electron temperature is enhanced to more than 9000 K during 
outbound orbit 1417.}
\end{center}
\end{figure}

Figure 4a shows the nightside electron temperature profiles with SZA during 16 
to 24 June 2015 (47 inbound orbits from 1381 to 1428). The electron temperature increases to 
$\sim$14000 K during orbit 1416. The increase in temperature can be observed in the  profiles  
obtained from the nightside legs of five consecutive inbound orbits from orbit 1415 to 1419. Figure 
4b shows the dayside electron temperature profiles with SZA during 16 to 24 June 2015 (47 outbound 
orbits from 1382 to 1429). The electron temperature increases to $\sim$11000 K during outbound of 
orbit 1417 on the dayside ionosphere.

\begin{figure}
\begin{center}
\includegraphics[width=1\textwidth]{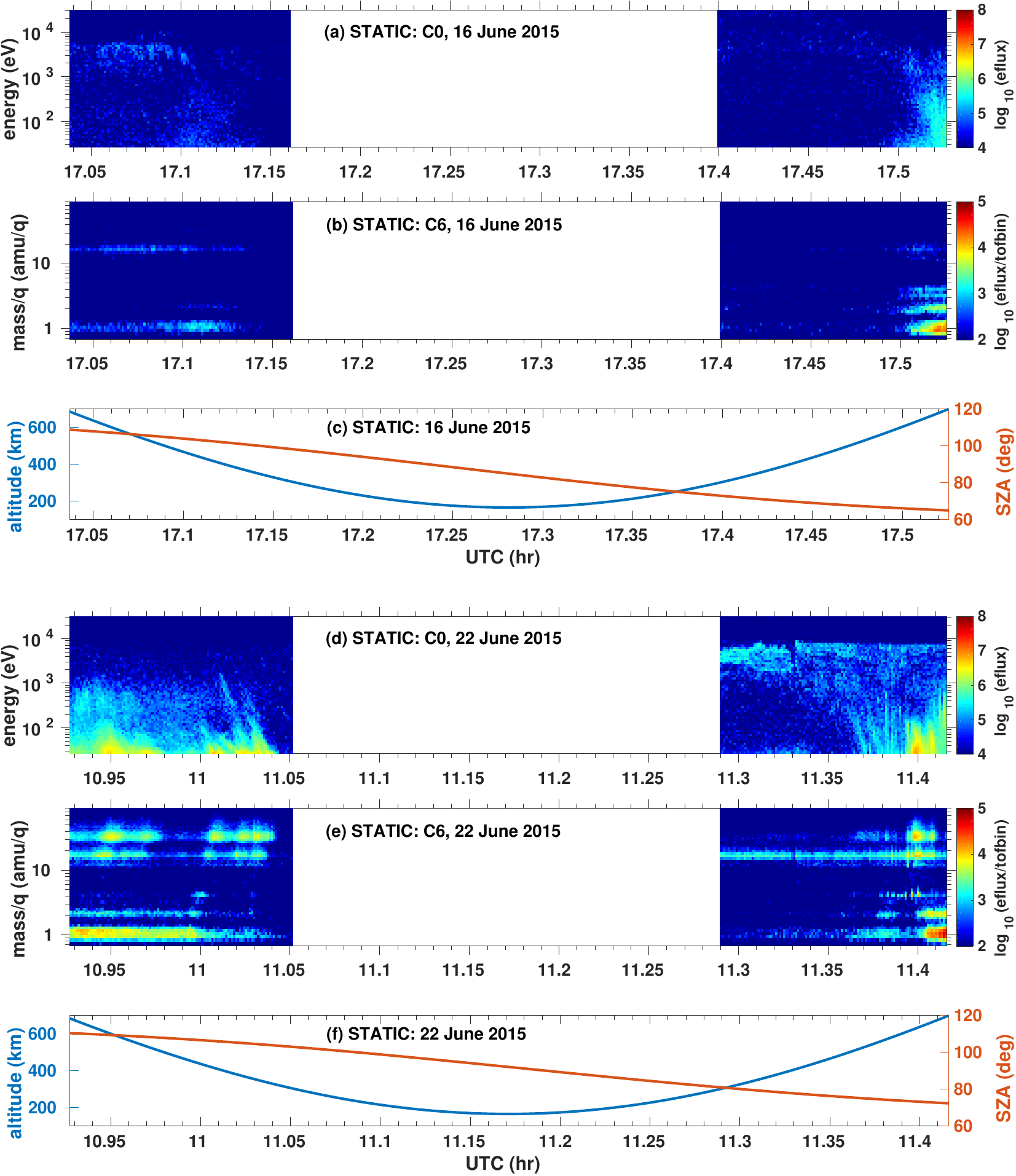}
\caption{STATIC energy-time spectrogram of omnidirectional ion energy flux (C0 mode) during (a) 
orbit 1384/1385, and (d) orbit 1415/1416, STATIC mass-time spectrogram of omnidirectional ion energy 
flux (C6 mode) during (b) orbit 1384/1385, and (e) orbit 1415/1416 (c, f) altitude and SZA during 
the MAVEN orbit leg corresponding to the STATIC observations. The eflux is expressed in 
units of differential energy flux (eV/[eV cm$^2$ sr s]). The white gaps on (a, b, d, 
e) corresponds to altitudes below 300 km.}
\end{center}
\end{figure}

\begin{figure}
\begin{center}
\includegraphics[width=1\textwidth]{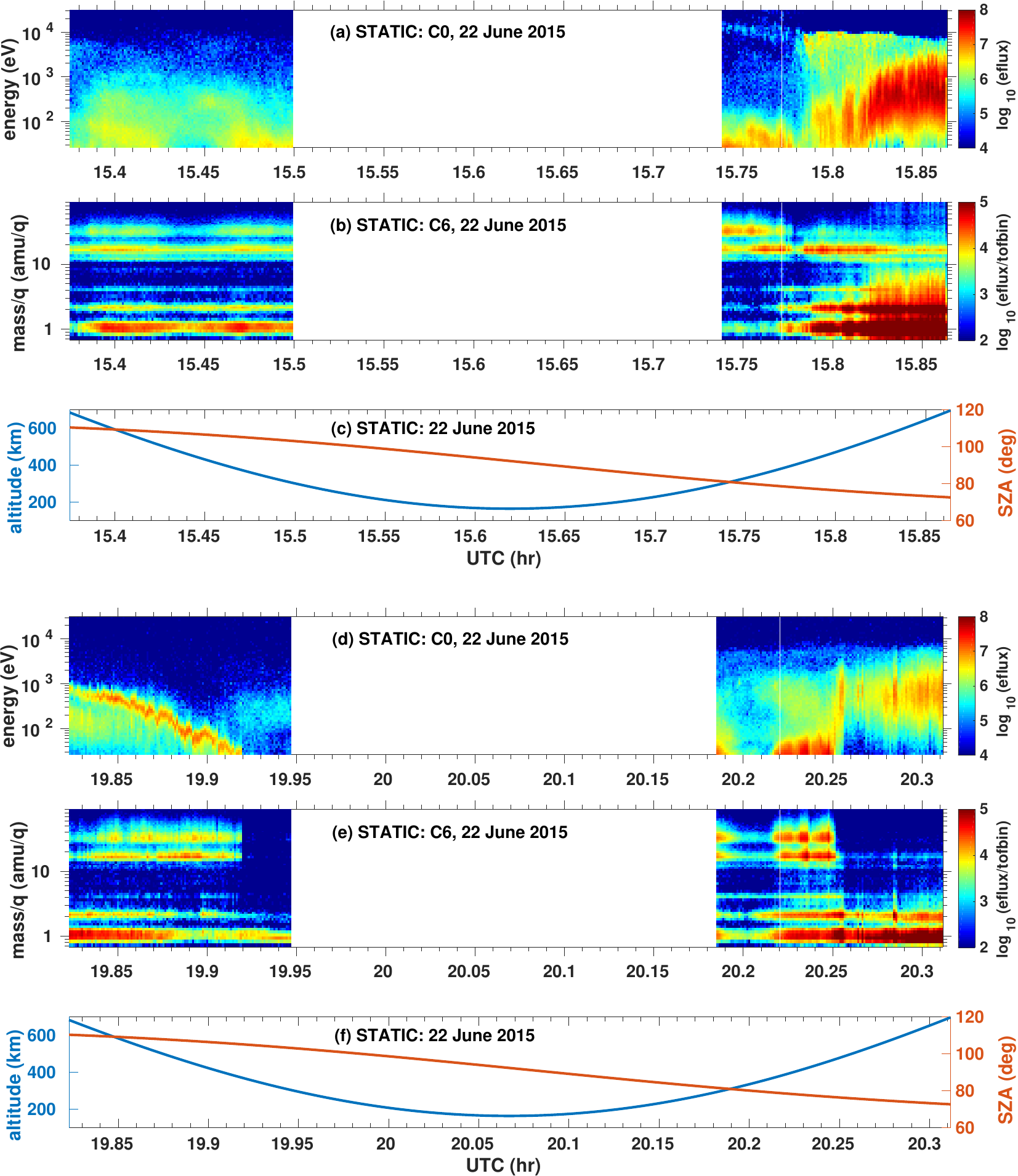}
\caption{STATIC energy-time spectrogram of omnidirectional ion energy flux (C0 mode) during (a) 
orbit 1416/1417, and (d) orbit 1417/1418, STATIC mass-time spectrogram of omnidirectional ion 
energy flux (C6 mode) during (b) orbit 1416/1417, and (e) orbit 1417/1418 (c, f) altitude and SZA 
during the MAVEN orbit leg corresponding to the STATIC observations. The eflux is expressed in 
units of differential energy flux (eV/[eV cm$^2$ sr s]). The white gaps on (a, b, d, 
e) corresponds to altitudes below 300 km.}
\end{center}
\end{figure}

\begin{figure}
\begin{center}
\includegraphics[width=1\textwidth]{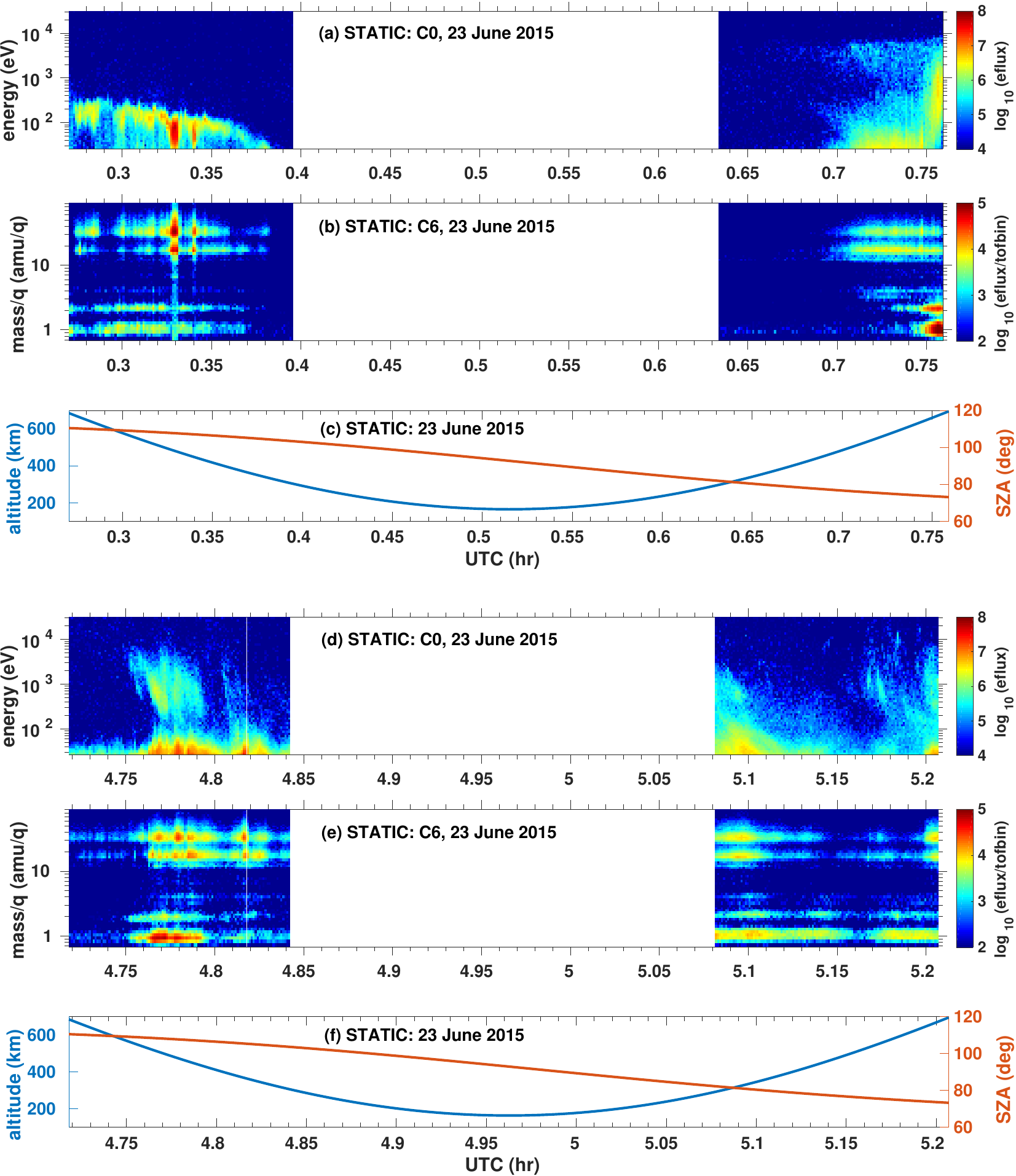}
\caption{STATIC energy-time spectrogram of omnidirectional ion energy flux (C0 mode) during (a) 
orbit 1418/1419, and (d) orbit 1419/1420, STATIC mass-time spectrogram of omnidirectional ion 
energy flux (C6 mode) during (b) orbit 1418/1419, and (e) orbit 1419/1420 (c, f) altitude and SZA 
during the MAVEN orbit leg corresponding to the STATIC observations. The eflux is expressed in 
units of differential energy flux (eV/[eV cm$^2$ sr s]). The white gaps on (a, b, d, 
e) corresponds to altitudes below 300 km.}
\end{center}
\end{figure}

Figures 5 to 7 shows the STATIC energy-time spectrogram and mass-time spectrogram of omnidirectional 
ion energy flux in C0 and C6 modes, respectively, during quiet (Figures 5a-5c) and disturbed 
(Figures 5d-5f, 6, 7) periods. The altitude and SZA during the MAVEN orbit legs corresponding to the 
STATIC observations are also shown. Figures 5a-5c shows the spectrogram during quiet time, before 
the arrival of CIR, and this corresponds to the periapsis of orbit inbound 1384/outbound 1385. We 
can see that there is no significant flux of heavy ions observed at higher altitudes. STATIC can 
measure suprathermal ions in the ionosphere and corona, and we can see a significant enhancement in 
flux of O$^+$ (amu/q=16) and O$_2^+$ (amu/q=32) ions as a consequence of the CIR arrival (Figures 
5d-5f, 6, 7). We can also observe the enhanced flux of penetrating solar wind protons (H$^+$, 
amu/q=1) into the ionospheric altitudes during CIR period (e.g. Figure 5b). The picked up heavy ions 
shown in the vicinity of Mars are of energies above 25 eV upto few keV (e.g. Figure 6a). The STATIC 
measured ion energy flux returns to the quiet time state afterwards (orbit number 1420/1421), with 
spectra very similar to that shown in Figures 5a-5c (not illustrated). This is in agreement with the 
LPW observations of electron density profiles on the nightside, where the profiles returns to quiet 
time state (orbit 1420 shown in Figure 3a).\\ 

It may be noted that STATIC was operating at energies above 25 eV during the observation period 
shown in this paper. Hence, the detected ions represents the suprathermal tail of ram ions, which 
are tailward accelerated by the enhanced solar wind. The observations above 300 km 
only are shown, so that the spacecraft potential is minimal and can be assumed to be small (that 
is, about -2 V). It may also be noted that the ion energy flux in the STATIC mass spectrograms are 
divided with number of TOF bins in each mass bin, since the different TOF bins are 
summed in order to maintain a constant mass independent of energy (J. P. McFadden, personal 
communication).\\ 

The  inbound orbits 1383, 1384 (quiet), 1416 and 1417 (disturbed), as well as outbound orbits  1384, 
1385 (quiet), 1417 and 1418 (disturbed) passed through strong crustal field regions in the southern 
hemisphere. During quiet periods, the topside ionospheric profiles over the crustal field region 
were more or less identical to those measured outside the crustal field region. Therefore, the 
observed topside depletions during the disturbed period can be  directly linked  with the enhanced 
solar wind conditions, rather than the presence of crustal magnetic fields. 



\section{Discussion} 
The CIR shock accelerated a significant number of particles at 1.5 AU, which are detected as 
enhanced flux of SEP ions \citep{Thampi2019}. These enhanced energetic particles and CIR-associated 
magnetic field compressions could significantly affect the topside ionosphere of Mars, and the 
topology of Mars\textendash solar wind interaction. The observed lower ionopause altitude 
($\sim$400 
km) during orbit 1417 is primarily due to the enhanced solar wind dynamic pressure, and the 
dayside ionosphere returns to the normal state in the declining speed region of the HSS, on the 
next orbit itself (Figure 3b). During orbit 1417 outbound, the electron density below 200 km is 
increased to $4\times 
10^4$ cm$^{-3}$, which is around 200\% increase in response to the topside day time compression, in 
comparison with quiet time mean. This is similar to the response during an 
interplanetary CME \citep{Thampi2018}. However, a striking feature in response to the CIR passage 
is that on the nightside, the topside ionosphere is significantly depleted for 5 
consecutive 
orbits, from orbit 1415 to orbit 1419 (Figure 3a). The electron temperatures derived from the LPW 
indicate increased temperature (above 9000 K) only during orbit 1417 on the dayside (Figure 4b), 
while, on the nightside the observed increase in temperature is higher (above 12000 K) and persists 
for more than one orbit (Figure 4a).\\

This is similar to the observations reported by \citet{Cravens1982} on the Venusian nightside 
ionosphere that the depleted electron densities are accompanied by enhancement in electron 
temperatures  along with large coherent horizontal magnetic fields. The solar wind dynamic pressures 
were observed to be larger than average conditions during the orbits when ionosphere is found to be 
depleted. The dayside ionopause was seen to be at a lower altitude when the solar wind dynamic 
pressure is large. Therefore, if the vertical extent of the dayside ionosphere is severely reduced, 
then the reservoir for the sustenance of nightside ionosphere can become smaller, and the supply of 
the ions would be curtailed \citep{Cravens1982}. In the present observations during CIR event, this 
is not the likely scenario for the extended depletion of topside ionosphere, because the dayside 
compression is observed only for a shorter duration. For the Venusian nightside ionosphere, 
\citet{Cravens1982} had suggested that the large coherent and horizontal magnetic field observed for 
depleted ionosphere might have (a) inhibited the downward diffusion, thus reducing the supply of 
ions to lower altitudes, and (b) inhibited the wake  electron precipitation, thus shutting down a 
significant source for nightside ionization.  In the case of Mars, these possibilities are not 
likely because of two reasons: (1) at lower altitudes, the differences of night time 
electron density compared to the quiet time are not significant. In the case of Mars, the downward 
diffusion is not the major source for sustaining the nightside ionosphere \citep{Fowler2015}. 
Therefore, the extent of nightside ionosphere need not be linked directly to the dayside ionopause 
altitude, and this is in concurrence with the LPW observations. (2) If the electron precipitation 
were significantly curtailed, the ionosphere in the lower altitudes also might have depleted 
significantly \citep{Fowler2015}, which is not the case in this event.\\

The Martian ionospheric ions are produced in the dayside and are picked up by the solar wind, 
accelerated and are swept away in the solar wind electric field direction. During CIRs, the mass 
loading of the solar wind plasma increases due to a deeper penetration of the interplanetary 
magnetic flux tubes into the ionosphere. The gyroradius of picked up O$^+$ ions before CIR passage 
(on 20 June 2015) is about $\sim$30100 km (corresponding to the average solar wind velocity of 300 
km sec$^{-1}$ and magnetic field strength of 1.6 nT). After CIR passage (on 22 June 2015), the 
gyroradius of O$^+$ pickup ions is around $\sim$8500 km (corresponding to the average solar wind 
velocity of 450 km sec$^{-1}$ and magnetic field strength of 8.7 nT). The ratio of gyroradii after 
and before CIR shock is about 0.28. The reduction in gyroradii of heavy ions may explain the 
enhanced STATIC ion energy fluxes at 
exospheric altitudes observed during the CIR. This may also explain the enhanced STATIC ion energy 
fluxes in the nightside close to the terminator above 300 km altitude.\\

These values of gyroradii are similar to those reported by \citet{Hara2011} inferred using MEX 
observations and ACE 
observations at $\sim$1 AU. Hence the results indicate that when the strength of 
the IMF  is enhanced due to the compressed IMF structure of the CIR, the gyroradii of the picked 
up planetary ions, such as O$^+$ becomes smaller (comparable to the Martian diameter) 
and therefore they can be found in the nightside ionosphere of Mars in the vicinity of ionospheric 
and exospheric altitudes. Using MEX data, \citet{Edberg2010} showed that the tailward flux of O$^+$ 
in the Martian ionosphere is significantly higher during CIR passage compared to quiet times. The 
photoionization of neutrals in the topside ionosphere produce ions which can be picked up by the 
sheath plasma and carried to the nightside \citep{Cravens1982}. This low density mass loaded 
ionosheath flow will also deplete the topside night time ionosphere.  Our observations of enhanced 
heavy ion flux in the nightside sector corroborates with this (Figures 5-7). The present study also 
indicate that these enhanced energetic heavy ions can significantly deplete the topside ionosphere, 
as evident from the topside electron densities in the nightside observed by the LPW instrument. 
Therefore, the present observations suggest that the enhanced pickup and sputtering in the nightside 
during CIR event is the major cause for the depleted ionization on the topside night time ionosphere 
of Mars. The feature of occurrence of the heavy ion precipitation during CIR passage is different 
from the conventional expectation of constant ion precipitation. Hence the dynamic solar wind 
conditions indeed play a major role in the efficiency of pickup and sputtering processes and hence 
the atmospheric escape at Mars.

\section{Summary and Conclusions}

The response of ionosphere of Mars to the passage of CIR of June 2015 is studied using observations 
from MAVEN mission. The CIR arrival was observed at Mars by enhancements in solar wind density, 
velocity, and dynamic pressure, and increased and fluctuating IMF. The CIR event was accompanied by 
solar energetic particles (that is, energetic electrons and ions). The LPW instrument onboard MAVEN 
provides the electron density and electron temperature of the Martian ionosphere. We observe that 
the dayside ionosphere is significantly compressed during the peak of solar wind dynamic pressure 
enhancement ($\sim$14 nPa). This response is similar to the dayside ionospheric response to an 
interplanetary CME. But in contrast, we observe that the nightside electron density depletes for a 
longer period of time. The electron temperatures are also enhanced during the period of thermal 
plasma depletion on the Martian nightside ionosphere. STATIC instrument onboard MAVEN observed 
enhanced fluxes of energetic protons and heavy ions (energies above 25 eV upto few keV) in the 
Martian 
exosphere during the impact of CIR. The planetary ions picked up by the solar wind are observed as 
enhanced flux of heavy ions in the vicinity of Mars at ionospheric altitudes. The gyroradius of 
O$^+$ pickup ions are observed to be relatively small during the ion 
precipitation. The observations indicate that the low density mass loaded tailward flow could be a 
major candidate for depleting the topside night time ionosphere. 

\acknowledgments
The work is supported by the Indian Space Research Organisation (ISRO). The LPW electron density 
(Level 2, version 2, revision 2) and electron temperature data (Level 2, version 3, revision 2), 
STATIC ion energy flux data (Level 2, version 2, revision 0), SWIA solar wind ion data (Level 2, 
version 1, revision 1), SWEA solar wind electron data (Level 2, version 4, revision 1), MAG IMF data 
(Level 2, version 1, revision 2), and SEP solar wind energetic particle flux data (Level 2, Version 
4, Revision 2) used in this work are taken from the Planetary Data System ({https://pds.nasa.gov/}). 
We gratefully acknowledge the MAVEN team for the data. We thank James P. McFadden, Space Sciences 
Laboratory, University of California, Berkeley for the helpful discussion on STATIC data. The 
WSA\textendash ENLIL+Cone simulations are taken from {http://helioweather.net/}, 
{https://iswa.ccmc.gsfc.nasa.gov/}. C. Krishnaprasad acknowledges the financial assistance provided 
by ISRO through a research fellowship.







\end{document}